\documentclass[a4paper,fleqn,usenatbib]{mnras}
\usepackage{amssymb,amsmath,graphicx,psfrag,mathtools}
\usepackage[T1]{fontenc} 
\usepackage{ae,aecompl} 
\usepackage{url}
\title[FRB 121102 Burst Pairs]{FRB 121102 Burst Pairs}
\author[J. I. Katz]{
J. I. Katz,$^{1}$\thanks{E-mail katz@wuphys.wustl.edu} 
\\
$^{1}$Department of Physics and McDonnell Center for the Space Sciences,
Washington University, St. Louis, Mo. 63130 USA 
}
\date{Accepted XXX.  Received YYY; in original form ZZZ} 
\pubyear{2017} 
\date{\today}
\begin{document} 
\label{firstpage} 
\pagerange{\pageref{firstpage}--\pageref{lastpage}} 
\maketitle 
\begin{abstract}
The repeating FRB 121102 emitted a pair of { apparently} discrete bursts
separated by 37 ms and another pair, 131 days later, separated by 34 ms,
during observations that detected bursts at a mean rate of $\sim 2 \times
10^{-4}$/s.  { Here I assume that these events are separate bursts rather
than multiple peaks from longer single bursts and consider their
implications.  They then are} inconsistent with Poissonian statistics.  The
measured burst intervals constrain any possible periodic modulation
underlying the highly episodic emission.  If more such short intervals are
measured a period may be determined or periodicity may be excluded.  Narrow
wandering beams are predicted to produce an excess of pairs (or higher
multiples) of bursts spaced at intervals much less than the mean interval.
\end{abstract}
\section{Introduction}
{ The repeating FRB 121102 is unique.  Its repetitions permitted accurate
localization and and identification with a star-forming dwarf galaxy at a
redshift $z = 0.193$ \citep{C17,T17,B17}.  Its redshift resolved all doubt
that (if it is representative of FRB, aside from its repetition rate, as it
appears to be) FRB are at ``cosmological'' distances, implying high emitted
instantaneous power, despite a small duty factor.  Repetition demonstrated
that FRB, unlike GRB, are not the product of catastrophic events.
Observations of repetitions over several years exacerbated the stringent
requirements on the energy available, and test theoretical models, such as
extreme pulsar pulses and SGR outbursts \citep{K16}.  Comparatively frequent
bursts offer the opportunity to obtain information, such as the distribution
of intervals between bursts, unavailable from FRB not observed to repeat.}

\citet{S17} at the Green Bank Telescope and \citet{H17} at Effelsberg have
observed pairs of radio bursts from the repeating FRB 121102 separated by
$37.3 \pm 0.3$ ms and $34.1 \pm 0.3$ ms, respectively (error estimates from
\citet{H17}, assumed the same for \citet{S17}, and propagated as
root-sum-of-squares).  Using the mean rate of burst detections of $\sim 2
\times 10^{-4}$/s in both studies, if Poissonian statistics applied only a
fraction $\lesssim 10^{-5}$ of bursts would be found in such close pairs.  The
detection of two such pairs in observations comprising (together) 25 bursts
is thus extraordinarily unlikely unless bursts are correlated on very short
time scales.  Many possible models are consistent with correlation.  No
single model is specifically indicated, but any successful model must admit
such short-time correlations.

{ In this paper I consider implications of the observations of these short
intervals between bursts.  { \citet{Sc16} show one burst (10) with either
two components separated by about 10 ms or a FWHM of about 10 ms; the
structure is frequency-dependent and the noise level is significant.  The
greatest FWHM shown by \citet{S17} is about 3 ms while the greatest FWHM
indicated by the Gaussian fits of \citet{H17} is about 5 ms.  The signals
shown in these latter two papers, with high S/N, have no indication of any
power in the $\sim 30$ ms between the two closely spaced peaks.  This makes
it unlikely that these peaks $\sim 35$ ms apart are substructure of a single
broad burst.  I therefore assume that the reported bursts are in fact
separate bursts.}

{ I first consider the hypothesis that SGR are analogous to giant pulsar
pulses \citep{K16} in which} the radiated power is drawn from the rotational
energy of a neutron star.  Bursts of FRB 121102 have been reported to have
fluxes as high as 0.8 Jy \citep{H17}, though this may be enhanced by
scintillation { or lensing \cite{Co17}}.  A conservative lower bound {
might} be 0.1 Jy.  This implies, assuming isotropic emission, a power of
$1.5 \times 10^{41}$ ergs/s.  If spindown power is converted to coherent GHz
radiation with efficiency $\epsilon$ the maximum spin period would be $120
\epsilon^{1/4} B_{15}^{1/2}$ ms.  \citet{C06} have found separations of
sub-pulses in the radio emission from an AXP of $\sim 0.1$ of the spin
period.  Unless $B_{15} \gtrsim 10 \epsilon^{-1/2}$, requiring an
unprecedentedly high $\epsilon$ (the frequency of giant pulses of the Crab
pulsar sharply decreases for $\epsilon \gtrsim 10^{-5}$ \citep{KSS10}) and
$B$ orders of magnitude greater than in any known magnetar, it is not
possible to explain the $\approx 35$ ms burst intervals as multiple phase
windows within a single longer rotation period { of a rotation-powered
FRB}.

SGR-like models, dissipating magnetostatic energy, are not bound by these
constraints, and \citet{H17} suggested that the short intervals may
represent multiple rotational phases of a single (slower) rotation.  Such
models are disfavored by the absence of a FRB in a fortuitous radio
observation during the extraordinary 2004 outburst of SGR 1806-20; its
Galactic distance of $\sim 15$ kpc would imply a brightness about 110 dB
greater than that of FRB at cosmological distances, more than compensating
for a 70 dB sidelobe supression at $35^\circ$ from the beam, yet none was
detected \citep{T16}.  { This argument depends on the unverified
assumption that FRB emit roughly isotropically; if strongly beamed, the
argument is vitiated.}

If the burst source has an underlying periodicity with extensive nulling,
like RRAT { Rotating Radio Transients \citep{M06}}, a template { (if
scaled up by many orders of magnitude in energy)} for a pulsar-based model
of FRB, then these observations constrain possible periods.  { If there
is a single rotational phase of emission} the period must be an integer
fraction of { all} the observed pair intervals.  The observation of more
than one such interval requires also that the period be an integer fraction
of the differences between each of the pair intervals, a strong constraint.
The observation of multiple pair intervals could either unambiguously
determine a millisecond period or exclude the possibility of such underlying
periodicity.
\section{The Period}
{ \citet{Sp16,H17} pointed out that the discovery of repeating pulses
offers an opportunity to determine if they are periodic, and their period
if so, even if (as in RRAT) pulses are observed very infrequently.  This
method becomes less effective if the periods are very much shorter than the
intervals between observed pulses, and may fail entirely if the period
varies so that the observed phase cannot be maintained between successive
observations.  This is likely if the FRB is produced by a very fast and
high-field neutron star that rapidly spins down, { if there are orbital
Doppler shifts or glitches, or if the emission region is in a surrounding
nebula and moves}.  Phase may be lost if the intervals between observations
$t_{int} \gtrsim \sqrt{1/{\dot \nu}} \sim \sqrt{T/\nu}$ where $T$ is the
spindown time.  For hypothetical $T \sim 100$ y and $\nu \sim 500$/s phase
may be maintained over intervals of 1--2 hours, comparable to the intervals
observed during periods of { apparent} activity \citep{Co17}, provided
irregular timing noise is small.  Phase may be maintained over longer
intervals if the spindown rate follows a simple function like a power law
that can be fitted.}

Suppose a strictly periodic underlying phenomenon in which an overwhelming
majority of possible pulses (more than 99.9999\% in the recent observations)
are nulled below the detection limit.  This (with less extreme nulling) is
the generally accepted model of RRAT.  It would also describe giant pulsar
pulses were there sufficiently high thresholds of detectability, and many
pulsars show more or less frequent nulling.  Pairs of pulses must be
separated by an integer multiple of the period, so that the period
\begin{equation}
\label{onepair}
P = {\Delta T_i \over n_i}
\end{equation}
for some positive integer $n_i$, where $\Delta T_i$ is the interval between
two bursts in the $i$-th pair.  In addition, for all pairs $(i,j)$
\begin{equation}
\label{twopairs}
P = {\Delta T_i - \Delta T_j \over k_{i,j}}
\end{equation}
for some positive integer $k_{i,j}$.  These resemble Diophantine equations,
but are complicated by the fact that the $\Delta T_i$ have errors of
measurement.

{ Measurement uncertainty limits the efficacy of period determination.}
{ A single pair of bursts with an interval $t_{int} \sim
5000\,$s, timed to an accuracy $\delta t \sim 0.3\,$ms, is consistent with
\begin{equation}
N_\nu \sim t_{int} \delta t\,\nu^2 \sim 4 \times 10^5
\end{equation}
distinct possible spin periods
of frequency $\nu = {\cal O} (500$/s).  As \citet{Sp16,H17} pointed out, the
extant data are not sufficient to determine a fast spin period, if there be
one, or to exclude its existence.  Each additional independently measured
interval of length $\sim 5000\,$s prunes the set of possible periods by a
factor $\sim \nu \delta t \sim 0.1$, so that many intervals with $t_{int}
\sim 5000\,$s must be measured to determine a unique fast period.}

For the { close} pairs observed by \citet{S17} and \citet{H17}, Eq.~\ref{twopairs} is
the strongest constraint, and implies that $P$ must be an integer fraction
of $3.2 \pm 0.4$ ms (propagating errors as root-sum-of-squares).  The
integer is unlikely to be more than about five because a neutron star has a
(somewhat uncertain) minimum rotational period of about 0.6 s.  Periods
permitted by the extant data are $3.2 \pm 0.4$ ms, $1.6 \pm 0.2$ ms, $1.07
\pm 0.13$ ms, $0.81 \pm 0.10$ ms, $0.65 \pm 0.08$ ms, $0.54 \pm 0.07$ ms
{\it etc.\/}  { If periodic at all, the period must be $\lesssim 3$ ms.}
Such short periods are also required to meet the energetic requirements of
pulsar models \citep{K16} provided they are not narrowly beamed and their
instantaneous radiated power does not exceed the spindown power (see,
however, \citet{K17a,K17b} for speculative alternatives).

{ These uncertainties correspond to phase lags of many cycles over a
$\sim 5000$ s interval, so that the longer intervals, however accurately
measured, cannot be used to select a valid, or exclude an invalid, short
period directly from those permitted by the 34.1 and 37.3 ms intervals.  It
may be possible to ``ladder up'' through a series of measured intervals
$\Delta T_n$, $n = 1, 2,\ldots$ satisfying
\begin{equation}
{\Delta T_{n+1} \over \Delta T_n} \lesssim {P \over 2 \delta t},
\end{equation}
where the $T_n$ can be differences between measured intervals.  For
periods $\sim 1$ ms this ratio is not large.  Suitable intervals have not
yet been measured.}
\section{Close Pairs}
If burst arrival times are described by Poissonian statistics with a mean
rate $\tau^{-1}$, then the {\it a priori\/} likelihood that an interval
between bursts is less than $T$ is $1 - e^{-T/\tau} \approx T/\tau$ (if
$T/\tau \ll 1$).  For the close pairs observed by \citet{S17,H17}, $\tau
\sim 5000\,$s and $T/\tau \sim 7 \times 10^{-6}$.  The {\it a posteriori\/}
choice of the observed intervals as the criterion $T$ introduces a bias that
invalidates the quantitative applicability of the {\it a priori\/}
likelihood, but it is still apparent that the process is far from
Poissonian.

The distribution of intervals may be consistent with a model in which a
narrow beam executes a random walk in direction, { whether the beam is
emitted by a neutron star or a black hole accretion disc \citep{K17a,K17c}}.
In such models the statistics of recurrence are non-Poissonian.  If the beam
once points to the observer, producing an observable burst, the interval
before the next burst is likely to be much less than its mean for Poissonian
statistics, the reciprocal of the mean burst rate.  This may be estimated
from the probability that a two dimensional random walk in angle (the space
of angular deviations is nearly Cartesian for small deviations) will return
to a particular direction in the interval $t$ to $t+dt$ after its previous
visit to that direction.  That probability density, for small deviations, is
$\propto t^{-1}$ (because the dispersion $\sigma \propto t^{-1/2}$ in each
of two orthogonal directions).  Hence the likelihood of a return within a
time $T$
\begin{equation}
\label{rw}
P(T) = {\int_{T_0}^T {dt \over t} \over \int_{T_0}^{T_{max}} {dt \over t}}
	= {\ln{T/T_0} \over \ln{(T_{max}/T_0)}},
\end{equation}
where $T_0$ is a lower cutoff corresponding to the time required for the
beam to wander its own width and $T_{max}$ an upper cutoff on the recurrence
time (in this model, corresponding to the time for the beam to return to a
direction to the observer following diffusion to the outer bounds of its
angular range).  Neither of these parameters is well known (the observed
pulse widths set an upper bound to $T_0$ but are broadened by scintillation,
imperfect de-dispersion and instrumental response), but the dependence of
$P(T)$ on them is only logarithmic.  For $T_0 = 1$ ms (a plausible upper
limit) and $T_{max} = \tau \sim 5000\,$s, $P(\text{50 ms}) \approx 0.25$.
This result should not be taken quantitatively, but indicates that in a
wandering beam model short recurrence times occur orders of magnitude
($10^3\text{--}10^4 \times$ for our parameters) more frequently than would
be indicated by Poissonian statistics.

{ The data are shown in Fig.~\ref{data}.  Each subfigure shows the
intervals reported in the indicated paper, binned in widths of $\sqrt{10}$
on a logarithmic scale.  The solid lines show the predictions of Poissonian
statistics with the mean burst rate observed during the period over which
the recurrences were observed.  This rate varies over times of hours, days
and longer (Table 1 of \citet{Sp16} and Table 2 of \citet{Sc16} indicate
periods of greater and lesser { apparent \citep{Co17}} activity like
those of SGR \citep{L87}) so the predictions are not quantitative, but
confirm the conclusion that the existence of repetition intervals $\sim 35$
ms is strong evidence against the Poissonian model even during periods of
greater activity (when nearly all the observed bursts occur and intervals
are measured).  The dotted lines show the predictions of the random walk
model (Eq.~\ref{rw}), with rolloffs allowing for the finite length of
continuous observations (intervals longer than the time from a burst to the
end of the observation cannot be observed).  This model is statistically
consistent with the observations of millisecond intervals, although there
may be a significant deficiency of intervals between 0.1 s and hundreds of
seconds that is not explained by the model.}
\begin{figure}
\centering
\includegraphics[width=0.99\columnwidth]{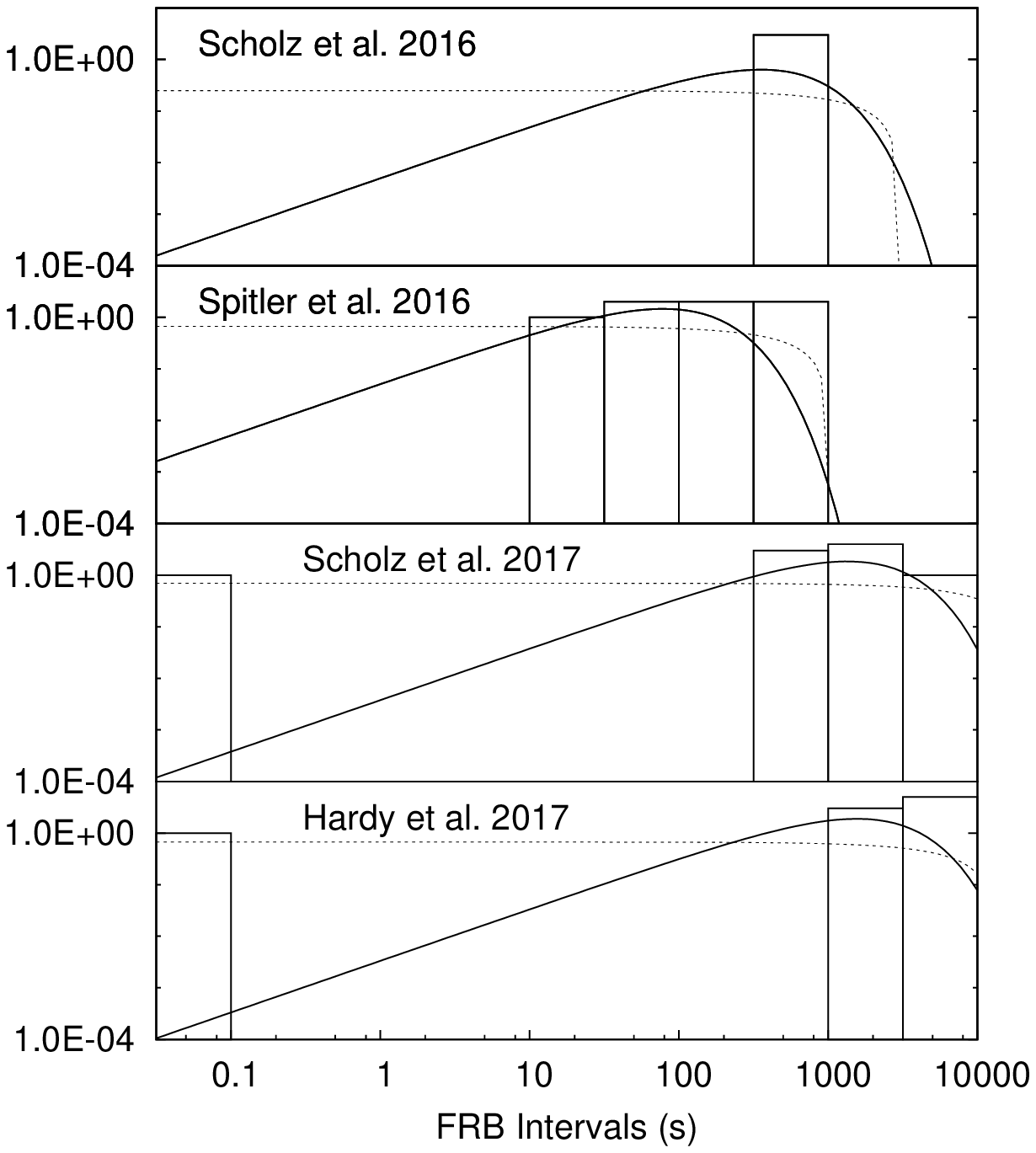}
\caption{\label{data}  Distributions of burst intervals in bins
$\sqrt{10}$ wide on a logarithmic scale, observed by \citet{Sc16} (Green
Bank Telescope, 2 GHz), \citet{Sp16} (Arecibo, 1.4 GHz), \citet{S17} (Green
Bank Telescope, 2 GHz with one interpolated Arecibo burst, January 12, 2017
data only) and \citet{H17} (Effelsberg, 1.4 GHz, data from January 16, 19,
25, 2017 only).  The shorter intervals observed by \citet{Sp16} at Arecibo
are consistent with a more sensitive telescope that can detect fainter
bursts unobservable elsewhere. The solid curves are predictions for
independent random events (Poissonian statistics).  The likelihood of
finding two intervals in the $10^{3/2}$--$10^2$ ms range out of 21 Green
Bank and Effelsberg intervals, assuming Poissonian statistics, is $\lesssim
10^{-5}$ (the Arecibo bursts \citep{Sp16} are not included in this estimate
because of their higher rate; there were no Arecibo intervals $< 10$ s).
The dotted curves are predictions for a narrow beam random walking in angle
and are consistent with the two sub-second recurrences.  They may not be
consistent with the gap between the very short and the longer intervals, but
the statistics are poor.}
\end{figure}
\section{Discussion}
The remarkable discoveries by \citet{S17,H17} of closely spaced pairs of
bursts from the repeater FRB 121102 offer important clues to FRB mechanisms.
If there is an underlying periodic clock, such as neutron star rotation,
{ with period short enough that the paris cannot be structure of single
bursts} then these pairs strongly constrain possible periods.  Because every
pair of closely spaced bursts provides an additional independent $\Delta
T_i$, the number of constraints provided by Eq.~\ref{twopairs} grows
quadratically with the number of pairs observed.  The discovery of one or
two more pairs with separations comparable to those recently observed could
either conclusively demonstrate the existence of a periodicity and determine
its period (and hence demonstrate the validity of pulsar-like models) or
demonstrate the absence of periodicity (and hence disprove such models
{ in which the emission region is stable in phase}).  

The observation of close pairs with highly non-Poissonian statistics
requires explanation.  Even in a periodic model, the observation of close
pairs requires that, apart from the periodicity, the source's activity be
correlated in time.  Many astronomical objects do have time-clustered
non-Poissonian statistics, with SGR as perhaps the most dramatic examples
\citep{L87}.  While these clearly distinguish periods of activity from less
active periods, they do not resemble the close pairs of FRB 121102.  Unlike
SGR and other episodically active objects (and phenomena outside astronomy
such as earthquake swarms), { a general period of enhanced activity is
not sufficient to explain} the close pairs of FRB 121102.  Perhaps they can
be explained, with statistical plausibility, if a collimated beam executes a
random walk in direction.  It returns to the observer more often shortly
after having pointed in that direction than long after, when it has wandered
far away.

\bsp 
\label{lastpage} 
\end{document}